%% file: arxiv.tex
\documentclass{article}

\PassOptionsToPackage{numbers, compress}{natbib}

\usepackage[preprint]{neurips_2026}

\usepackage[utf8]{inputenc} 
\usepackage[T1]{fontenc}    
\PassOptionsToPackage{hyphens}{url}           
\usepackage{hyperref}       
\usepackage{booktabs}       
\usepackage{amsfonts}       
\usepackage{nicefrac}       
\usepackage{microtype}      
\usepackage[table]{xcolor}  
\usepackage{colortbl}
\usepackage{natbib}
\usepackage{pifont}
\usepackage{multirow}
\usepackage{subcaption}
\usepackage{tcolorbox}
\tcbuselibrary{breakable}
\input{utils}

\newcommand{\methodname}{AgentCARD\xspace}

\title{Specialize Roles, Mix Deployments: Pushing the Cost--Accuracy Frontier of LLM Agent Teams}

\author{%
  Yinsicheng Jiang\textsuperscript{1} \And
  Liang Cheng\textsuperscript{1} \And 
  Yeqi Huang\textsuperscript{1} \And
  Yufan Zhao\textsuperscript{1} \And
  Zhan Lu\textsuperscript{1} \AND
  Li Dong\textsuperscript{2} \And
  Wenda Li\textsuperscript{1} \And
  Edoardo Ponti\textsuperscript{1} \And
  Luo Mai\textsuperscript{1} \And
  \vspace{-10pt} \\
  \textsuperscript{1}University of Edinburgh \quad
  \textsuperscript{2}Microsoft Research \quad
}

\begin{document}

\maketitle

\begin{abstract}
LLM agents are increasingly deployed as multi-role teams, where tasks are divided across specialized roles such as planner, executor, and verifier. In these systems, cost and accuracy are no longer properties of a single model: they depend on which model fills each role and where it is hosted, including API, self-hosted, and hybrid deployment. Existing agentic benchmarks typically evaluate fixed models or fixed agent configurations, and therefore offer limited guidance for cost--accuracy-optimal deployment. We introduce \methodname, a role-aware benchmark suite for evaluating LLM agent teams across role assignment and deployment mode. \methodname combines a role-decomposed evaluation harness, a unified API/self-hosted cost model, Pareto-frontier analysis, and a Shapley-based diagnostic for identifying role bottlenecks. Our evaluation shows that heterogeneous teams consistently occupy the cost--accuracy frontier. They improve accuracy by up to \textbf{44\%} over cost-equivalent homogeneous teams, or match the strongest homogeneous team at up to $\textbf{12}\times$ lower per-task cost through hybrid deployment. We further find that the best role assignment is domain-dependent: some domains are planner-bottlenecked, while others are executor-bottlenecked. Finally, \methodname extends beyond planner--executor teams to workflows with additional roles such as verification, and supports continual evaluation as new domains and team structures emerge. Our code is released at: \url{https://github.com/Auto-CAP/AgentCAP}.
\end{abstract}

\input{secs/1_introduction}
\input{secs/2_background}
\input{secs/3_implementation}
\input{secs/4_accuracy_matrix}
\input{secs/5_deployment}
\input{secs/6_role_criticality}
\input{secs/7_insights}

\bibliography{mybib}
\bibliographystyle{unsrtnat}

\clearpage
\appendix
\input{secs/appendix}

\end{document}

%% file: utils.tex
\usepackage{glossaries}
\makeglossaries
\usepackage{tikz}

\newcommand{\tinyskip}{\vspace{3pt}}
\newcommand{\mypar}[1]{\tinyskip\noindent\textbf{#1.}\xspace}

\usepackage{mathtools}
\usepackage{xspace}

\makeatletter
\DeclareRobustCommand\onedot{\futurelet\@let@token\@onedot}
\def\@onedot{\ifx\@let@token.\else.\null\fi\xspace}

\makeatother


%% file: secs/1_introduction.tex
\section{Introduction}
\label{sec:intro}

LLM applications are increasingly organized as \emph{role-decomposed agent teams}, where different agents specialize in roles such as planning, execution, verification, and tool use. This pattern now appears in production coding agents~\citep{openai_codex2025, cursor2026, opencode2025}, open-source multi-agent frameworks~\citep{wu2023autogen, google_adk2026}, and large-scale orchestration systems~\citep{kimi_agent_swarm2026}. Yet the practical value of role separation remains unclear. Recent industry evidence shows both large gains and large costs: Anthropic's multi-agent research system reports a $+90.2\%$ accuracy improvement over its single-agent counterpart while spending roughly $15\times$ more tokens~\citep{anthropic_multi_agent2025}, whereas Cognition~\citep{cognition_dont_build2024} cautions that naive role replication can multiply cost without proportional gains. This creates a practical evaluation question: \emph{when does assigning different models to different roles improve task success enough to justify the added cost?}

Recent single-agent baselines further expose the limits of fixed-agent evaluation. On open-domain tool use, MCP-Atlas~\citep{scale_mcp_atlas2026} reports that Claude-Opus-4.6 and GPT-5.4 pass only about 50\% of tasks, with many failures caused by refusals, missing tool-use opportunities, and tool-call loops that expand context to $600\text{K}+$ tokens. These failures suggest that open-domain tool use is not only a model-capability problem, but also an orchestration problem: the agent must maintain the task framing, decide the next action, and manage tool interactions over long contexts. The plan-execute paradigm~\citep{wang2023plansolve, yao2023react, shen2023hugginggpt, shinn2023reflexion, kim2024llmcompiler} addresses this problem by separating high-level decomposition from multi-turn tool execution. Its extensions, such as plan-verify-execute~\citep{erdogan2025planandact, copilotcli_critic2026, cursor_planmode2026}, add further specialized roles. However, this trend also changes what a benchmark must measure: the relevant question is no longer simply which model is best, but which model should fill which role, on which task domain, and under which deployment mode.

Benchmarking role-decomposed agents is therefore fundamentally different from benchmarking single agents. Once a task is run by an agent team, accuracy and cost are no longer properties of an individual model; they are properties of a \emph{role assignment}, a \emph{deployment mode}, and a \emph{task domain}. This creates three challenges. \textbf{(i)}~\emph{Pairwise accuracy}: the planner shapes the executor's action space, turn count, and failure modes, so the same model can be strong in one role and weak in another. \textbf{(ii)}~\emph{Asymmetric cost}: planners and executors have different token profiles, and API, self-hosted, and hybrid deployments produce different dollar-per-task tradeoffs. Per-token API prices alone vary by $5\text{--}8\times$ across frontier models~\citep{openrouter2025}; self-hosting open-weight models with optimized hardware and inference systems can further change the serving cost~\citep{narayanan2024melange, moecap2024}. \textbf{(iii)}~\emph{Domain-dependent bottlenecks}: upgrading the planner and upgrading the executor can have opposite effects across domains. Existing agentic benchmarks typically report task success for a fixed model or fixed agent configuration. They do not jointly measure task success, cost per task, and role bottlenecks across role assignments and deployment modes.

\begin{figure}
    \centering
    \begin{minipage}{0.5\linewidth}
        \centering
        \includegraphics[
        width=\linewidth,
        ]{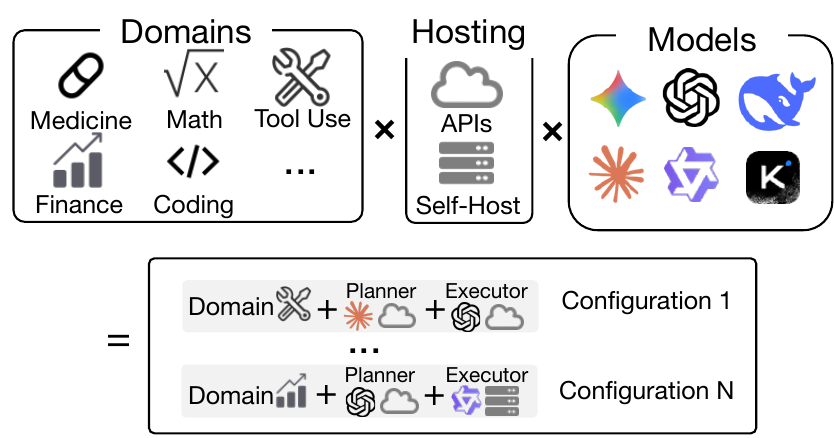}
    \end{minipage}
    \begin{minipage}{0.49\linewidth}
        \centering
        \includegraphics[width=\linewidth]{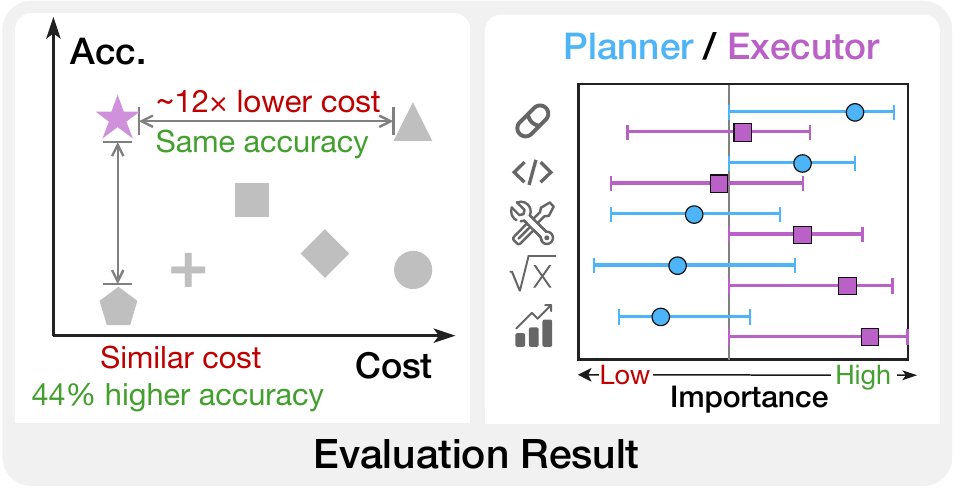}
    \end{minipage}
    \caption{Overview of AgentCARD. Agent teams, such as planner--executor and planner--verifier--executor teams, are evaluated across task domains, role assignments, and deployment modes. AgentCARD reports task accuracy, dollar-per-task cost, and role-criticality diagnostics on a unified cost--accuracy frontier.}
    \label{fig:overview}
    \vspace{-0.6cm}
\end{figure}

We introduce \methodname, a role-aware benchmarking suite for evaluating the cost--accuracy frontier of agent teams across \textbf{C}ost, \textbf{A}ccuracy, \textbf{R}ole assignment, and \textbf{D}eployment mode. Rather than introducing another static task collection, \methodname turns existing agentic benchmarks into role-decomposed evaluations. It provides an evaluation harness for agent teams (e.g., planner--executor and planner--verifier--executor), a unified API/self-hosted cost model, a Shapley-based role-criticality diagnostic, and an automated pipeline that refreshes the leaderboard as new models appear. We instantiate \methodname on five existing benchmark datasets covering open-domain tool use, finance, medical reasoning, coding, and mathematics. \methodname has three main contributions.

\mypar{Role-aware accuracy evaluation}
We evaluate all 16 planner--executor assignments among four API models, together with self-hosted alternatives, across five domains. We find that team accuracy is direction-aware: which model fills which role matters. Heterogeneous teams improve accuracy by up to $44\%$ when models are assigned to their stronger roles, while reversing the same pair can reduce accuracy by $36\%$.

\mypar{Deployment-aware cost--accuracy frontier}
We compute directly comparable per-task costs for API, self-hosted, and hybrid teams. Across domains, the Pareto frontier is consistently occupied by heterogeneous teams. A cheaper model can reach the frontier when assigned to a suitable role and paired with a stronger model in the other role. As a result, heterogeneous teams can match the strongest homogeneous team at $\mathbf{1.4}\times$--$\mathbf{12}\times$ lower per-task cost.

\mypar{Role-criticality diagnostics}
We propose a Shapley-based role-criticality diagnostic that identifies the bottleneck role per domain, reducing model search from every planner-executor pair to a single role, an order-of-magnitude saving at scale. Planner versus executor capability returns differ by up to $40\%$, so without the diagnostic, a budget upgrade routinely lands on the role that contributes nearly nothing. The verdict is preserved across the three deployment modes we test, and a plan-verify-execute pilot extends the framework to three-agent teams with up to a $9.4\%$ accuracy lift over the strongest two-agent baseline at comparable per-task cost. A held-out forecast validates that calibration-set diagnostics transfer.

%% file: secs/2_background.tex
\section{Background and Motivation}
\label{sec:background}
Agent-team evaluation differs from conventional agent evaluation because the unit of evaluation changes. In a standard agent benchmark, a model or a fixed agent configuration is evaluated on a task suite, with task success rate and sometimes cost as the result. This suffices when the deployment decision is which model or agent system to use, but not for role-decomposed teams. In a planner--executor team, a practitioner must decide which model fills each role (including whether the same model fills both) and whether each role is served through an API, self-hosted inference, or a hybrid of the two. The benchmarked object is therefore a team configuration, not a single model.

These agent teaming configuration choices change benchmark outcomes by orders of magnitude rather than by a small constant factor. Prior work reports up to $50\%$ accuracy differences from swapping which model fills the planner role in a planner-executor team~\citep{pear2026}; on our domains, the same model pair can give $+44\%$ synergy in one role direction and $-36\%$ in the reverse direction, an $80\%$ swing from a single configuration choice (Section~\ref{sec:accuracy_matrix}). Per-task cost varies by $5$--$35\times$ at matched accuracy across deployment modes: API token prices alone span $5$--$8\times$ across frontier providers, and self-hosting an open-weight executor on amortised hardware further reduces cost by $2$--$5\times$ relative to its API equivalent~\citep{handshake_anyscale_2024,databricks_llama3_2024}. 

Recent agentic benchmarks---AgentBench~\citep{kim2025agentbench}, $\tau$-bench~\citep{yao2024taubench}, MultiAgentBench~\citep{zhu2025multiagentbench}, TheAgentCompany~\citep{xu2024agentcompany}, MAESTRO~\citep{ma2026maestro}, HAL~\citep{kapoor2025hal}, and PEAR~\citep{pear2026}---have advanced LLM agent evaluation, but each fixes either the team or the cost model: PEAR sweeps planner--executor pairs yet reports only accuracy, while others report token usage or API cost only for a fixed configuration. None constructs a cost--accuracy frontier over both role assignments and deployment modes. Existing benchmarks therefore leave three deployment-critical questions unanswered: which model should fill each role, which team gives the best cost--accuracy tradeoff, and which role limits team performance.

\mypar{Fixed configurations hide role assignment}
Existing benchmarks evaluate either single-agent systems~\citep{kim2025agentbench, yao2024taubench, kapoor2025hal} or multi-agent systems with predetermined roles~\citep{zhu2025multiagentbench, xu2024agentcompany, ma2026maestro}. This answers which system performs best under a fixed setup, but not which assignment of models to roles is best within an agent team. For planner--executor agents, this distinction is central. The planner determines how the goal is decomposed and what intermediate steps the executor receives, while the executor determines whether those steps can be grounded through tool calls and multi-turn interaction. As a result, the same pair of models can behave differently depending on which model plans and which model executes. A benchmark that fixes role assignment cannot expose this asymmetry.

\mypar{Single-point cost hides the cost--accuracy frontier}
The benchmarks that report cost~\citep{kim2025agentbench, ma2026maestro, kapoor2025hal} usually attach a cost number to one model or one agent configuration. This is not enough for agent-team deployment. Different roles have different token profiles, so replacing the planner and replacing the executor can have very different effects on dollar-per-task cost. Deployment mode further changes the comparison: API models are priced per input and output token, while self-hosted models depend on GPU amortisation, utilization, and inference throughput. A benchmark useful for deployment should therefore compare configurations on a common cost axis and report the cost--accuracy frontier. This allows practitioners to identify the cheapest team that reaches a target accuracy, or the most accurate team within a target budget.

\mypar{Aggregate team scores hide role bottlenecks}
A single team-level accuracy number does not explain which role limits performance. This matters because the main lever in agent-team deployment is deciding where to spend model capability. If failures are mainly caused by poor planning, a stronger executor may increase cost without improving success rate. If failures are mainly caused by weak execution, a stronger planner may have little effect. Existing benchmarks do not provide this role-level diagnostic. \methodname addresses this gap by decomposing the contribution of role-specific model upgrades through a Shapley-based role-criticality analysis (Section~\ref{sec:role_criticality}); held-out validation of the resulting recommendations is reported in Appendix~\ref{subsec:eval_forecast}.

%% file: secs/3_implementation.tex
\section{Designing a Role-Aware Agent-Team Benchmark}
\label{sec:impl}

Section~\ref{sec:background} shows that benchmarking agent teams requires evaluating role assignment, cost--accuracy tradeoffs, and role bottlenecks. \methodname operationalizes these requirements by treating a \emph{team configuration} as the unit of evaluation. A configuration specifies the model and deployment mode (API, self-hosted, or hybrid) for each role in the team, together with the task domain; we instantiate it on plan-execute (two roles) and plan-verify-execute (three roles). For each configuration, \methodname records accuracy, role-separated token usage, and per-task cost. This lets the benchmark compare teams on a common cost-accuracy axis and decompose their accuracy into per-role contributions.

\mypar{Domains}
\methodname is instantiated on leading state-of-the-art benchmarks, creating a strong coverage in agentic AI domains. Each domain exposes a different balance between planning and execution. MCP-Atlas~\citep{scale_mcp_atlas2026} evaluates open-domain tool use, where agents must choose tools and recover from failed calls. MedAgentBench~\citep{jiang2025medagentbench} evaluates clinical API use, where success depends on selecting and executing appropriate information-gathering actions. FinanceBench~\citep{islam2023financebench} evaluates financial document QA, where planning identifies relevant evidence and execution grounds the answer. IMO-AnswerBench~\citep{luong2025imobench} evaluates math with Python tools, where planning decomposes the problem and execution implements and checks computations. SWEBench-Lite~\citep{jimenez2024swebench} evaluates software engineering patches, where planning localizes the change and execution edits, tests, and iterates on code. Together, these domains test whether role assignment and role bottlenecks vary across task families.

\mypar{Models and role assignments}
\methodname evaluates homogeneous and heterogeneous teams across API, self-hosted, and hybrid deployment modes. We use four API models---GPT-5.4, Claude-Opus-4.6, MiniMax-M2.7, and GLM-5.1---and four self-hosted models---Qwen3.5-4B, Qwen3.5-27B, GPT-OSS-20B, and GPT-OSS-120B. The API models span different price points, from MiniMax-M2.7 at \$0.40/\$1.60 to Claude-Opus-4.6 at \$5.00/\$25.00 per million input/output tokens (April 2026). The self-hosted models are served with vLLM 0.19.0 on a H100 cluster and priced using the cost model in Section~\ref{subsec:cost_model}.

For each domain, we evaluate all $4{\times}4$ API planner--executor pairs, API/self-hosted hybrid teams, and within-family open-weight pairs such as Qwen3.5-4B/Qwen3.5-27B and GPT-OSS-20B/GPT-OSS-120B. This matrix makes role assignment observable: it shows whether a model is better used as planner, executor, or both, and whether reversing the same model pair changes accuracy or cost.

\mypar{Unified harness}
All configurations run through a unified OpenCode-based harness~\citep{opencode2025, ohmyopencode2026}. The harness uses one tool registry and one plan-execute prompt scaffold, while routing planner and executor calls to independently configured models. This keeps differences attributable to the team configuration rather than to benchmark-specific scaffolding.

The harness covers dataset loading, plan generation, multi-turn execution, tool invocation, evaluation, and logging through a unified LLM client that supports API and self-hosted models with shared retry and token-accounting (input, output, cached) across configurations. Given a new API endpoint or self-hosted checkpoint, the pipeline reruns the planner--executor matrix and refreshes the cost--accuracy frontier and role-criticality diagnostics.

\mypar{Cost model}
\label{subsec:cost_model}
To compare API, self-hosted, and hybrid teams, \methodname uses a role-decomposed per-task cost model, $C_{\text{task}} = \sum_{r} C_r$, where $r$ indexes the team's roles (planner, executors, etc.) and $C_r$ is each role's per-task contribution. Each $C_r$ is computed by one of two rules, depending on how role $r$ is served.

For API-served roles, $C_r$ sums published token charges over all calls role $r$ issues during a task: $C_r = p_{\text{in}} N_{\text{in}}^{\text{fresh}} + p_{\text{cache}} N_{\text{in}}^{\text{hit}} + p_{\text{out}} N_{\text{out}}$, where $(p_{\text{in}}, p_{\text{cache}}, p_{\text{out}})$ are the model's published input, cached-input, and output prices and $(N_{\text{in}}^{\text{fresh}}, N_{\text{in}}^{\text{hit}}, N_{\text{out}})$ are the corresponding token counts aggregated over the role's calls. For self-hosted roles, $C_r$ is the GPU-hour price for role $r$'s reserved capacity, amortised over tasks completed during the benchmark run: $C_r = N_{\text{GPU},r}\times p_{\text{GPU/hr}}\times T_{\text{wall}} / (3600\times N_{\text{tasks}})$, where $T_{\text{wall}}$ is the wall-clock duration over which role $r$'s GPUs are held and $N_{\text{tasks}}$ is the number of team-level tasks completed in that window. Hybrid teams apply each rule to the corresponding role, yielding one comparable dollar-per-task value.

The GPU-hour price $p_{\text{GPU/hr}}$ is set to the Vast.ai H100 rate of \$1.87/hr~\citep{vastai_pricing_2026}; 
$T_{\text{wall}}$ and $N_{\text{tasks}}$ are taken from each benchmark's actual run. The auto-pipeline supports re-computing per-task cost at any user-specified parallelism level using the throughput surface in Appendix~\ref{app:throughput}. Reported costs cover LLM serving only; tool execution and orchestration overhead are $<5\%$ of the LLM cost and are excluded.

\mypar{Benchmark use cases}
\methodname supports three deployment use cases. First, the per-domain cost--accuracy frontier supports \emph{team selection}: choosing the cheapest team that reaches a target accuracy, or the most accurate team within a budget (Section~\ref{sec:deployment}). Second, role-criticality diagnostics support \emph{role upgrading}: identifying whether planner or executor quality is the bottleneck in each domain (Section~\ref{sec:role_criticality}). Third, the unified cost model supports \emph{deployment-mode selection}: comparing API, self-hosted, and hybrid teams under the same cost axis. Together, these use cases guide both current deployment and future migration as new API and open-weight models appear.

%% file: secs/4_accuracy_matrix.tex
\section{Pairwise Synergy of Plan-Execute Teams}
\label{sec:accuracy_matrix}
\begin{tcolorbox}{%
\textbf{Research Question 1}: Does pairing two different models in different roles produce genuine accuracy gains over the corresponding single-model teams for the same budget?
}
\end{tcolorbox}

\begin{figure}[t!]
    \centering
\includegraphics[width=\linewidth]{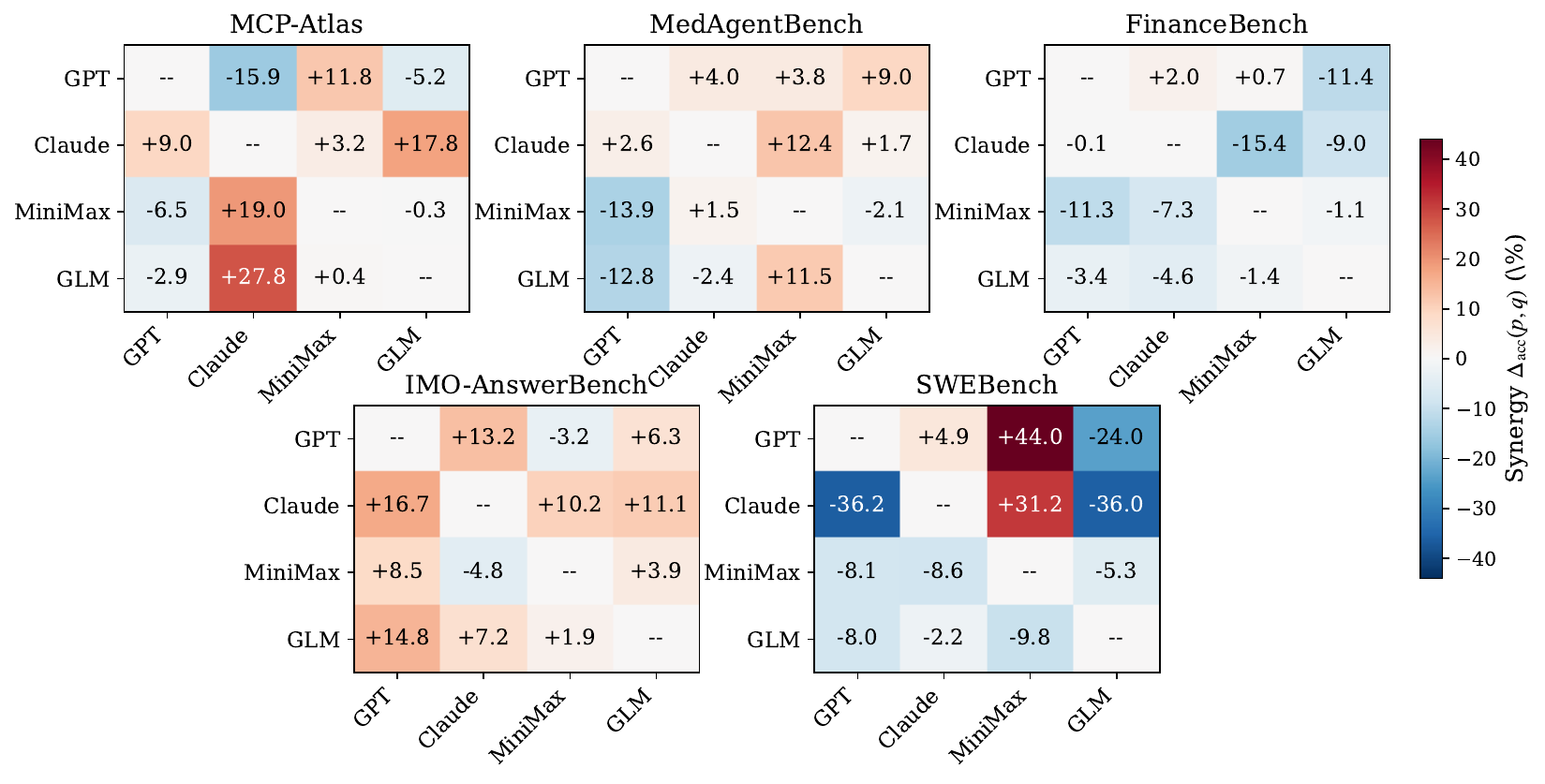}
    \caption{Pairwise synergy $\Delta_{\mathrm{acc}}(p,q)$ in \%. Lower triangle: planner $p$ with executor $q$; upper triangle: reversed pair. Warmer cells indicate positive synergy, colder cells negative synergy.}
    \label{fig:synergy}
\end{figure}

We characterize when and why heterogeneous teams are beneficial by evaluating planner-by-executor accuracy matrices across the five benchmarks listed in Section~\ref{sec:impl} (full matrices in Appendix~\ref{app:acc_other}) for each pair $(p, q)$ of API models. Then, we compute a cost-adjusted synergy that compares each heterogeneous team against \emph{budget-mixing} the two corresponding homogeneous teams $(p,p)$ and $(q,q)$. Concretely, denote by $\mathrm{acc}(p,q)$ and $\mathrm{cost}(p,q)$ the accuracy and per-task cost of the team with planner $p$ and executor $q$. The two homogeneous teams define a line in cost--accuracy space; we evaluate this line at $\mathrm{cost}(p,q)$ to obtain the baseline accuracy
  \[
    w \;=\; \mathrm{clip}\!\left(\frac{\mathrm{cost}(p,q)-\mathrm{cost}(p,p)}{\mathrm{cost}(q,q)-\mathrm{cost}(p,p)},\,0,\,1\right),
    \qquad
    \mathrm{acc}_{\mathrm{base}}(p,q) \;=\; (1-w)\,\mathrm{acc}(p,p) +
  w\,\mathrm{acc}(q,q),
  \]
  and the synergy is reported as the signed gap
  $\Delta_{\mathrm{acc}}(p,q) = \mathrm{acc}(p,q) -
  \mathrm{acc}_{\mathrm{base}}(p,q)$.
  A positive value indicates a real synergy: the heterogeneous team yields more
  accuracy than \emph{any} budget allocation between the two homogeneous teams
  at the same expected cost. We avoid the alternative of comparing $\mathrm{acc}/\mathrm{cost}$ ratios across teams because cost is unbounded upwards, whereas accuracy is bounded in $[0,100]$, so the cheapest team would almost always win regardless of accuracy; the
  convex-combination baseline removes this artefact by anchoring the comparison to a cost-matched alternative on the homogeneous line.

  Results are shown in Figure~\ref{fig:synergy}:
  The lower-triangular cells display the
  direction-specific $\Delta_{\mathrm{acc}}(p,q)$ for the configuration with planner $p$ and executor $q$; the upper-triangular cells display the reverse.

\mypar{Pairwise synergy is benchmark-dependent and direction-aware}
SWEBench is the most polarised domain: GPT-5.4 planning with MiniMax-M2.7 executing gains of $+44$ over budget-mixed homogeneous teams, while Claude-Opus-4.6 planning with GLM-5.1 executing loses $-36$. Heterogeneous teams gain up to $+27.8$ on MCP-Atlas, $+16.7$ on IMO-AnswerBench, and $+12.4$ on MedAgentBench. On the other hand, FinanceBench displays a flat surface near zero, because the four API models score within a tight accuracy band on this task family, and any heterogeneous pair tracks the budget line.

\mypar{Two conditions decide whether heterogeneous pairing pays off}
First, the two models must differ enough in domain skills that the homogeneous baseline line spans a non-trivial accuracy range; otherwise, a budget-matched mixture is hard to beat (as on FinanceBench). Second, the direction must match each model's role-specific advantage on that domain; otherwise, reversing the pair erases or inverts the gain (as in SWEBench's $+44$ versus $-8.1$ direction asymmetry). When both conditions hold, picking the right role for each model matters more than picking strong models, foreshadowing the per-domain bottleneck analysis of Section~\ref{sec:role_criticality}.

\begin{figure*}[t!]
    \centering
    \includegraphics[width=\linewidth]{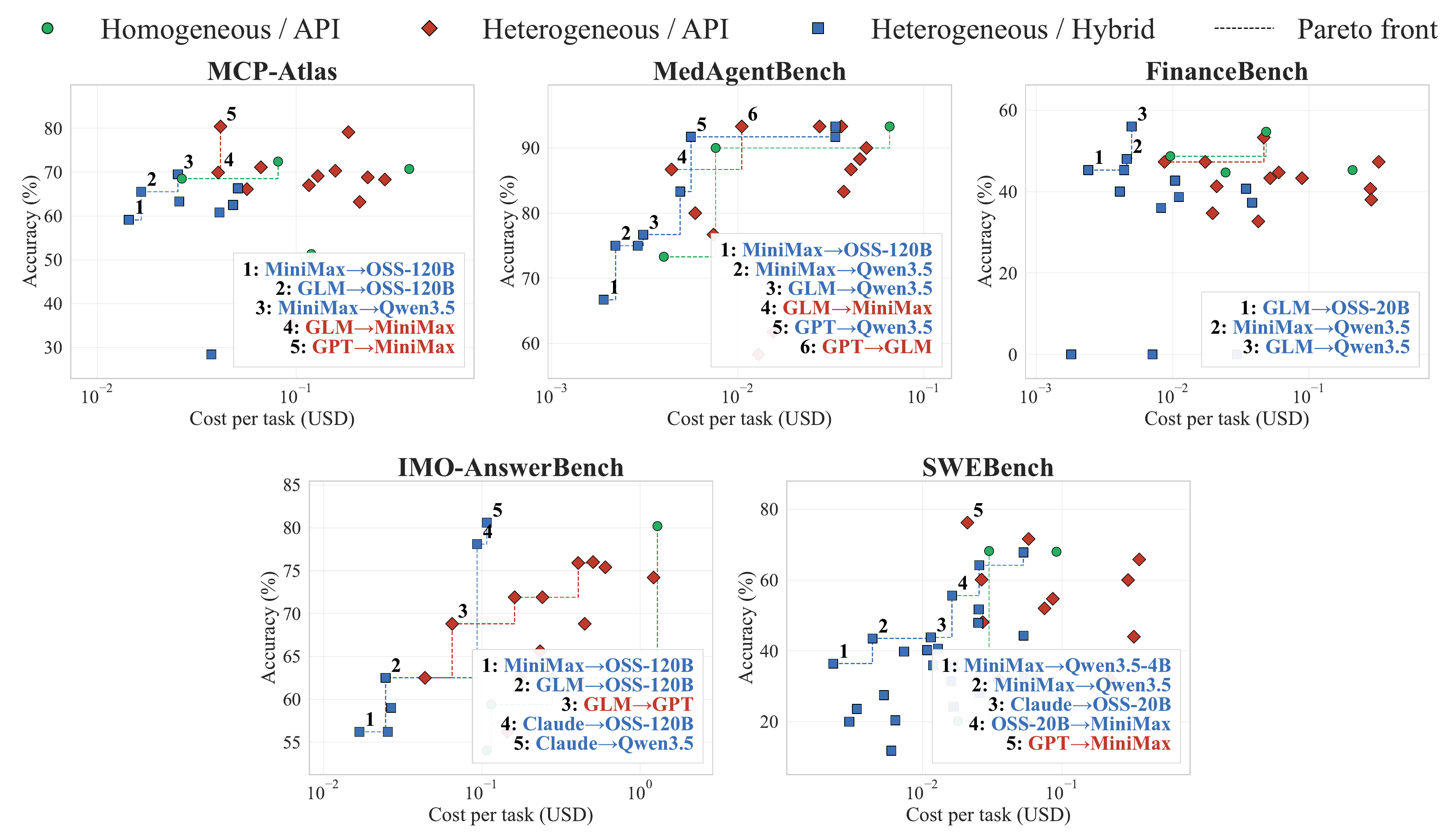}
    \caption{Cost--accuracy scatter plot for every (planner $\rightarrow$ executor) team across the five benchmarks. Each point is one team; color and shape encode the type of team (Homogeneous API, Heterogeneous API, or Heterogeneous hybrid) and the corresponding Pareto frontier. Numbered dots indicate teams lying on the global Pareto frontier.}
    \label{fig:pareto}
    \vspace{-0.4cm}
\end{figure*}

%% file: secs/5_deployment.tex
\section{Cost-Accuracy Modeling and Deployment-Aware Frontier}
\label{sec:method}\label{sec:deployment}
\begin{tcolorbox}{%
\textbf{Research Question 2}: Does varying the mode of deployment of each model in a pair (API, self-hosted, or hybrid) affect the Pareto frontier for a given domain?
}
\end{tcolorbox}

To map the cost--accuracy trade-off across LLM agent teams with different deployment modes (both API or hybrids where the planner is API and the executor is self-hosted), we plot every team on a shared cost--accuracy plot in Figure~\ref{fig:pareto}. 
Per-benchmark accuracy and cost per task for the four highest-accuracy heterogeneous teams and the four API self-play teams are tabulated in Appendix~\ref{app:cost_cap}. The frontier reported here is over the evaluated configurations; the auto-pipeline supports adding new candidates without modifying the harness.

\mypar{Heterogeneous role assignments occupy the frontier}
Across the five benchmarks, the Pareto frontier is occupied by heterogeneous role assignments rather than by homogeneous self-play teams. Two measurements explain the gap. \emph{On the cost axis}, the executor reads roughly $200\times$ more input tokens than the planner emits (mostly tool observations from the multi-turn loop), so the executor consumes a median of $6\times$ more dollars per task than the planner (Appendix~\ref{app:role_asymmetry}); routing execution to a cheaper role-competent model reduces total cost. \emph{On the accuracy axis}, the per-role best model differs from the per-task best model in 4 of the 5 benchmarks (Section~\ref{sec:accuracy_matrix} matrices), so pairing each role with the model that scores highest on that role lifts team accuracy beyond any single-model team. Multiple heterogeneous teams typically sit within bootstrap CI of the per-benchmark frontier (Appendix~\ref{app:acc_other}), giving practitioners cost-equivalent alternatives among top-accuracy configurations.

\mypar{Cheap models reach the frontier with the right partner}
A cheap model, open-weight or low-cost API, reaches the frontier when its partnered role is strong, regardless of which side it occupies. With the cheap model as planner, MiniMax-M2.7 paired with GPT-5.4 execution reaches $48.1\%$ on SWEBench ($2.4\times$ its self-play baseline), and Qwen3.5-4B paired with Qwen3.5-27B execution reaches $60.0\%$ on FinanceBench, the cheapest top-accuracy team. With the cheap model as executor, Claude-Opus-4.6 paired with Qwen3.5-27B execution reaches $80.6\%$ on IMO-AnswerBench at $12\times$ lower cost than Claude self-play, and $93.3\%$ on MedAgentBench at $2\times$ lower cost. The same pattern holds symmetrically on MCP-Atlas: removing the strong partner from either role drops accuracy by more than swapping the cheap model alone.

\mypar{Hybrid teams enter the frontier at a fraction of API cost}
Hybrid teams that mix API and self-hosted models reach the cost--accuracy frontier across the tool-use and reasoning benchmarks we test, e.g., Qwen3.5-27B as executor reaches up to 69.5\% on MCP-Atlas, 93.3\% on MedAgentBench, 80.6\% on IMO-AnswerBench with Claude-Opus-4.6 planning, and 60.0\% on FinanceBench with Qwen3.5-4B planning, all at $5$--$35\times$ lower per-output-token cost than the best all-API team. Combined with the $6\times$ role-asymmetric cost concentration above, mixing API and self-hosted deployments captures most of the savings while preserving accuracy on critical decisions.

%% file: secs/6_role_criticality.tex
\section{Domain-Specific Role Criticality}
\label{sec:role_criticality}
\label{subsec:bottleneck_swap}

\begin{figure}[t]
	\centering
	\includegraphics[width=\linewidth]{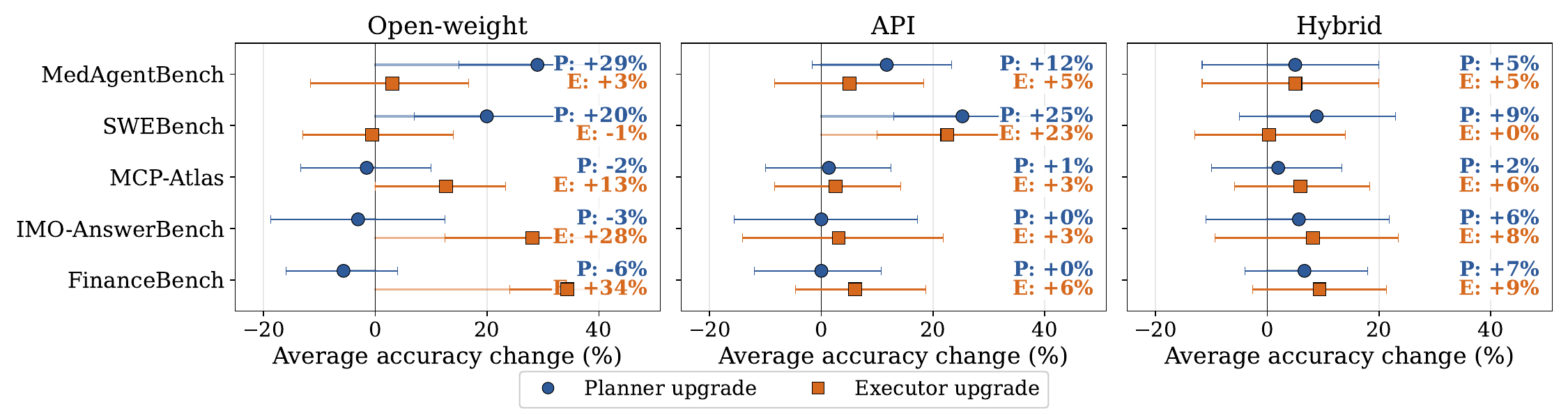}
	\caption{Shapley decomposition of role criticality under three deployment modes of weak-to-strong upgrade. Blue circles mark the planner upgrade $\phi_P$ and orange squares the executor upgrade $\phi_E$; error bars are 95\% bootstrap confidence intervals over tasks. Benchmarks are sorted by the planner-versus-executor swing difference averaged over the three modes.}
	\label{fig:shapley_role}
\end{figure}

\begin{tcolorbox}{%
\textbf{Research Question 3}: In which domains is planning or execution more critical?
}
\end{tcolorbox}

To diagnose which role gates accuracy on each domain, we Shapley-decompose the accuracy change from a controlled \emph{weak-to-strong upgrade} in each role. Formally, for a given pair of planner candidates $\{p_W, p_S\}$ and executor candidates $\{e_W, e_S\}$ where $W$ is the weaker member of each pair and $S$ is the stronger, we evaluate all four (planner, executor) combinations on the five domains and write $a_{xy} := \mathrm{acc}(p_x, e_y)$ for $x, y \in \{W, S\}$ as the per-task accuracy. The average accuracy change attributable to upgrading each role is the Shapley value
\begin{equation} \label{eq:shapley}
        \phi_P \;=\; \tfrac{1}{2}(a_{SW} - a_{WW}) + \tfrac{1}{2}(a_{SS} - a_{WS}),
    \qquad
    \phi_E \;=\; \tfrac{1}{2}(a_{WS} - a_{WW}) + \tfrac{1}{2}(a_{SS} - a_{SW}),
\end{equation}
  satisfying efficiency $\phi_P+\phi_E = a_{SS} - a_{WW}$. The factor $\tfrac{1}{2}$ averages the marginal contribution of each upgrade over the two possible orders in which the role upgrades could be applied. As an example, using Qwen3.5-4B as $W$ and Qwen3.5-27B as $S$ gives $a_{WW}{=}28.7$, $a_{SW}{=}20.0$, $a_{WS}{=}60.0$, and $a_{SS}{=}57.3$ on FinanceBench. Substituting these values in Equation~\ref{eq:shapley} yields $\phi_P=-5.7$ and $\phi_E=+34.3$, identifying the executor as the bottleneck. In general, the larger the Shapley value, the more critical a role in that domain. 
  
  We instantiate the upgrade in three deployment modes shown in Figure~\ref{fig:shapley_role}, which sorts the five benchmarks from planner-critical (top) to executor-critical (bottom). The \textbf{Open-weight} mode draws both roles from a single self-hosted family at two sizes (Qwen3.5-4B vs 27B, or GPT-OSS-20B vs 120B). The \textbf{API} mode contrasts MiniMax-M2.7 against GLM-5.1. The \textbf{Hybrid} mode crosses an API upgrade in one role with an open-weight upgrade in the other. Each $(W,S)$ pair has a stable weak-strong identity: open-weight pairs share a model family with $W$ the smaller and $S$ the larger member by parameter count (Qwen3.5-4B$\to$27B, GPT-OSS-20B$\to$120B), and the API pair MiniMax-M2.7$\to$GLM-5.1 has $W$ self-play below $S$ on all five benchmarks (Appendix~\ref{app:role_criticality}).

\mypar{The bottleneck role differs across domains}
MedAgentBench is planner-critical ($\phi_P{=}29.0\%$ versus $\phi_E{=}3.0\%$ in the open-weight mode), consistent with its task structure of selecting one correct API endpoint per question. FinanceBench ($\phi_E{=}34.3\%$), IMO-AnswerBench ($\phi_E{=}28.1\%$), and MCP-Atlas ($\phi_E{=}12.6\%$) are executor-critical, matching their long tool-call or document-extraction trajectories. SWEBench shifts with deployment scale: at open-weight only the planner upgrade helps ($\phi_P{=}20.0\%$, $\phi_E{=}-0.5\%$), but at the API frontier both roles contribute substantially ($\phi_P{=}25.3\%$, $\phi_E{=}22.6\%$). Across the five domains, no single role dominates universally.

\mypar{The bottleneck verdict is preserved across deployment modes}
Across all three panels of Figure~\ref{fig:shapley_role}, executor-critical benchmarks (FinanceBench, IMO-AnswerBench, MCP-Atlas) keep $\phi_E \geq \phi_P$ and planner-critical MedAgentBench keeps $\phi_P \geq \phi_E$; magnitudes shrink in API and Hybrid as both models approach frontier capability, and the hybrid MedAgentBench cell ties at $5.0\%$, but no mode reverses the verdict. The verdict is also family-stable: on SWEBench, both within-family swaps give $\phi_P > \phi_E$ (Qwen3.5-4B$\to$27B: $19.9\%$ vs $-0.5\%$; GPT-OSS-20B$\to$120B: $12.8\%$ vs $5.0\%$). A practitioner can therefore identify the bottleneck role from any one mode and apply the same upgrade strategy under a different deployment budget.

\section{Extension to Three Agent Roles: Planner, Executor, and Verifier}
\label{subsec:plan_verify}

\begin{tcolorbox}{%
\textbf{Research Question 4}: Do heterogeneous teams bring further benefits with $n\geq3$ roles?
}
\end{tcolorbox}

To address this question, we instantiate a third role on top of the plan--execute pipeline~\citep{erdogan2025planandact}: a verifier that audits the planner's output before the executor runs it. This plan-verify-execute structure has reached production in critic-style coding agents~\citep{copilotcli_critic2026, cursor_planmode2026}. The planner produces a plan; the verifier critiques it along three axes (completeness, executability, step dependencies); the executor runs the approved plan. If the verifier flags a concrete defect, the planner revises the plan based on the verifier's feedback. This loop repeats up to two times. Our verifier prompt follows the iterative-critic style of Self-Refine~\citep{madaan2023selfrefine} and the evaluator-optimizer agent pattern~\citep{anthropic_multi_agent2025}. Because LLMs asked to self-correct tend to flag too many false issues~\citep{huang2024selfcorrect}, the verifier defaults to approving the plan unless it finds a concrete defect.

Section~\ref{sec:accuracy_matrix} identifies MiniMax-M2.7 as the strongest executor on MCP-Atlas across API planners. We fix it as executor and sweep planner and verifier across GLM-5.1 and GPT-5.4, the two strongest paired API planners; for context we also evaluate single-model three-agent self-play. Table~\ref{tab:plan_verify} reports the results. The pairwise-synergy and Shapley role-criticality metrics introduced in Sections~\ref{sec:accuracy_matrix} and \ref{sec:role_criticality} extend to teams of any number of roles; the formal generalisation and computational scaling are derived in Appendix~\ref{app:scaling}.

\begin{table}[t]
\centering
\caption{Plan-verify-execute on MCP-Atlas. \textbf{Top block:} MiniMax-M2.7 fixed as executor; planner and verifier swept. 
\textbf{Bottom block:} single-model three-agent teams (one model in all three roles). 
Best team in bold, second-best underlined, in terms of accuracy and \$/task.
}
\label{tab:plan_verify}
\begin{tabular}{llllcc}
\toprule
Team size & Planner & Verifier & Executor & Accuracy ($\uparrow$) & \$/task ($\downarrow$) \\
\midrule
2 & GLM-5.1       & GLM-5.1 & MiniMax-M2.7  & 74.5 & 0.043 \\
3 & GLM-5.1       & GPT-5.4       & MiniMax-M2.7  & \textbf{83.8} & \underline{0.039} \\
3 & GPT-5.4       & GLM-5.1      & MiniMax-M2.7  & 75.3 & 0.050 \\
2 & GPT-5.4       & GPT-5.4      & MiniMax-M2.7  & \underline{80.4} & 0.070 \\
\midrule
1 & GLM-5.1       & GLM-5.1         & GLM-5.1       & 71.6 & 0.075 \\
1 & GPT-5.4       & GPT-5.4         & GPT-5.4       & 68.7 & 0.126 \\
1 & MiniMax-M2.7  & MiniMax-M2.7    & MiniMax-M2.7  & 72.4 & \textbf{0.029} \\
\bottomrule
\end{tabular}
\end{table}

\mypar{Three heterogeneous agents form the strongest configuration}
The strongest team places three different models in the three roles, GLM-5.1 planner, GPT-5.4 verifier, and MiniMax-M2.7 executor, and reaches $83.8\%$ on MCP-Atlas, exceeding the strongest two-agent baseline (GPT-5.4 planner with MiniMax-M2.7 executor at $80.4\%$) at lower per-task cost: verification audits a short plan rather than generating one, so the GPT-5.4 verifier is cheaper than placing GPT-5.4 in the planner role. Single-model three-agent teams cluster at $68$--$73\%$; adding agents without role specialization pays the cost of three calls but recovers only the accuracy of one. The accuracy lift therefore comes from cross-model specialisation rather than from adding a third agent role.

\mypar{Verifier value is bounded above by its planning competence on the same task}
The only configuration that lifts accuracy meaningfully above the strongest two-agent baseline uses a verifier (GPT-5.4) outperforming the planner (GLM-5.1) in the planner role (Section~\ref{sec:accuracy_matrix}: $80.4\%$ vs $69.9\%$). The other configurations either hold (GPT-5.4 self-verification at $80.4\%$), lift marginally (GLM-5.1 self-verification, $+4.6$ above $69.9\%$), or drop ($-5.1\%$ when GPT-5.4 plans and GLM-5.1 verifies), consistent with the documented limit of intrinsic LLM self-correction~\citep{huang2024selfcorrect}. A third agent earns its overhead only when the verifier could plan better than the planner itself, extending the two-agent role-fit pattern of Section~\ref{sec:deployment} to three roles. The same pattern holds on FinanceBench, where adding Qwen3.5-27B as verifier between the MiniMax-M2.7 planner and the GLM-5.1 executor lifts accuracy from $53.3\%$ to $62.7\%$ at comparable per-task cost (\$0.05 vs \$0.06).

\mypar{Scaling to hundreds of agents}
Beyond the three-role pipeline, production deployments~\citep{kimi_agent_swarm2026} spawn up to a hundred sub-agents in parallel, typically by replicating one model to fan out a single role and aggregating the results. \methodname's per-role findings carry over to this regime by construction: a model that ranks high on a role in a small team is the natural candidate to fill that role at scale. We leave the empirical study of fan-out aggregation effects (e.g., majority voting across replicated executors and joint within-role replication with cross-role specialization) to future work.

%% file: secs/7_insights.tex
\section{Conclusions and Takeaway Messages}
\label{sec:insights}

\methodname derives key insights that highlight both the potential and the design considerations of multi-role (e.g., plan--execute) agent teams.

\mypar{Heterogeneous teams reach the cost-accuracy frontier}
Across our five benchmarks, the Pareto frontier is occupied by heterogeneous role assignments rather than by homogeneous self-play teams (Section~\ref{sec:deployment}), with pairwise synergy gains up to $+44\%$ when each model fills its better role. The per-role best model differs from the per-task best model in 4 of the 5 benchmarks, so role-aware pairing yields cost-accuracy points that no single-model team reaches.

\mypar{Agent teams enable role-aware mixing of API and self-hosted models}
The two roles have asymmetric cost profiles: the executor reads $200\times$ more input tokens and consumes a median of $6\times$ more dollars per task than the planner (Section~\ref{sec:deployment}, grounded in Appendix~\ref{app:role_asymmetry}). Hybrid teams that mix API and self-hosted models reach the cost-accuracy frontier on the tool-use and reasoning benchmarks at $5$--$35\times$ lower per-output-token cost than the best all-API team, while preserving accuracy on critical decisions.

\mypar{The bottleneck role is domain-dependent and transfers across deployment modes}
The Shapley decomposition (Section~\ref{sec:role_criticality}) identifies MedAgentBench as planner-critical ($\phi_P{=}29.0\%$ versus $\phi_E{=}3.0\%$) and FinanceBench, IMO-AnswerBench, MCP-Atlas as executor-critical, with the bottleneck verdict preserved across open-weight, API, and hybrid deployment modes. Agent-team configuration therefore depends on the target task family rather than on a single leaderboard winner.

\mypar{Routing agent teams calls for joint evaluation across role, deployment, and cost dimensions}
The four dimensions we identify in Section~\ref{sec:background} (role assignment, cost-accuracy frontier, hybrid deployment, role decomposition) jointly determine which team to deploy, yet existing benchmarks evaluate at most a subset of them. A good agent-team router therefore combines three signals: the domain (which team sits on the frontier, Section~\ref{sec:deployment}), the bottleneck role (where capability spend matters most, Section~\ref{sec:role_criticality}), and per-role strengths (planner-best and executor-best are typically different, Section~\ref{sec:accuracy_matrix}). \methodname's auto-pipeline (Section~\ref{sec:impl}) keeps these signals current as the model pool evolves.

%% file: secs/appendix.tex
\section{Self-Hosted Throughput Profiles}
\label{app:throughput}

We profile Qwen3.5-27B and GPT-OSS-120B at TP${\in}\{2, 4, 8\}$ on H100~SXM~80GB with vLLM. Table~\ref{tab:tp_scaling} reports per-million-output-token cost at the headline operating point used in Section~\ref{sec:deployment}, at the \$1.87/hr GPU rate of Section~\ref{subsec:cost_model}.

TP${=}4$ is throughput-optimal for both models. Going to TP${=}8$ hurts Qwen3.5-27B by $3.7\times$ as tensor-parallel communication outpaces the added compute; GPT-OSS-120B is less penalized but its best point is also at TP${=}4$.

\begin{table}[!h]
\centering
\caption{Self-hosted \$/M output token at the headline operating point on $4{\times}$H100 SXM (\$1.87/hr per GPU). TP${=}4$ is throughput-optimal for both models. Prefill \$/M is within 5\% of decode and omitted. $T_{\text{dec}}$ is the aggregate server-level decode throughput measured at runtime with vLLM.}
\label{tab:tp_scaling}
\small
\begin{tabular}{lccc}
\toprule
                      & \textbf{TP$=$2}  & \textbf{TP$=$4}          & \textbf{TP$=$8}  \\
\midrule
Qwen3.5-27B (\$/M out)      & 0.82            & \textbf{0.46}            & 1.72             \\
\;\; $T_{\text{dec}}$ & 2562 tok/s      & 4565 tok/s               & 2442 tok/s       \\
GPT-OSS-120B (\$/M out)      & 0.48            & \textbf{0.29}            & 0.50             \\
\;\; $T_{\text{dec}}$ & 4375 tok/s      & 7241 tok/s               & 8400 tok/s       \\
\bottomrule
\end{tabular}
\end{table}

For the single-user on-premise case, we also profile Qwen3.5-4B and GPT-OSS-20B on a single RTX~5090, reaching \$0.11 and \$0.08 per million output tokens at a reference \$0.55/hr GPU rate.

\section{Per-Benchmark Results and Diagnostics}

\subsection{Accuracy Matrices for Other Benchmarks}
\label{app:acc_other}

Tables~\ref{tab:mcpatlas_main}, \ref{tab:swebench_main}, \ref{tab:med_matrix}, \ref{tab:fin_matrix}, and \ref{tab:imo_matrix} give the full planner--executor accuracy matrices for the five benchmarks. Across all five tables, rows index the planner, columns the executor; \textbf{bold} marks the best executor(s) per row, and \underline{underline} marks the best overall (note that multiple combinations are within each other's standard deviation). API models are accessed through OpenRouter (Claude/MM/GLM) or directly (GPT); self-hosted runs use $4{\times}$H100 at the per-model throughput-optimal TP from Appendix~\ref{app:throughput}.

\begin{table}[!h]
	\centering
	\caption{MCP-Atlas plan--execute accuracy (\%).}
	\label{tab:mcpatlas_main}
	\resizebox{0.98\textwidth}{!}{%
		\begin{tabular}{l|cccc|cc}
			\toprule
			\textbf{Planner / Exec} & \textbf{GPT-5.4} & \textbf{Claude-Opus-4.6} & \textbf{MiniMax-M2.7} & \textbf{GLM-5.1} & \textbf{Qwen3.5-27B} & \textbf{GPT-OSS-120B} \\
			\midrule
			GPT-5.4          & 70.7 $\pm$ 2.8 & 35.4 $\pm$ 2.0 & \textbf{\underline{80.4 $\pm$ 2.4}} & 67.0 $\pm$ 2.3 & 62.5 $\pm$ 3.0 & 63.3 $\pm$ 2.9 \\
			Claude-Opus-4.6  & 68.8 $\pm$ 3.2 & 51.3 $\pm$ 2.1 & 66.1 $\pm$ 2.5 & \textbf{69.1 $\pm$ 2.0} & 66.3 $\pm$ 2.3 & 60.8 $\pm$ 2.7 \\
			MiniMax-M2.7     & 63.2 $\pm$ 2.0 & \textbf{70.3 $\pm$ 2.3} & 68.5 $\pm$ 2.8 & \textbf{71.1 $\pm$ 2.7} & 69.5 $\pm$ 2.3 & 59.1 $\pm$ 2.8 \\
			GLM-5.1          & 68.3 $\pm$ 3.1 & \textbf{\underline{79.1 $\pm$ 2.0}} & 69.9 $\pm$ 3.0 & 72.4 $\pm$ 2.9 & 28.4 $\pm$ 2.4 & 65.5 $\pm$ 2.2 \\
			\bottomrule
		\end{tabular}%
	}
\end{table}

\mypar{Additional models on MCP-Atlas}
Kimi-K2.5, DeepSeek-v4-pro, and Hunyuan-3-preview were tested separately. Kimi-K2.5 self-play reaches 63.0\%; paired as planner with GPT-5.4, Claude-Opus-4.6, and MiniMax-M2.7 executors, it scores 57.5\%, 66.0\%, and 61.9\%, respectively; paired as executor under the same three planners, it scores 60.2\%, 63.4\%, and 58.3\%. DeepSeek-v4-pro planning Hunyuan-3-preview hits 65.8\%, dropping to 57.6\% with a MiniMax-M2.7 executor; the reverse direction (Hunyuan-3-preview planner, DeepSeek-v4-pro executor) collapses to 31.1\%.

\begin{table}[!h]
	\centering
	\caption{SWEBench plan--execute accuracy (\%). The strongest team (GPT-5.4 planning, MiniMax-M2.7 execution at 76.2\%) clears the strongest homogeneous team (GPT-5.4 self-play, 68.2\%) by 8.0 points. Three small open-weight planners (Qwen3.5-4B, GPT-OSS-20B, GPT-OSS-120B) all reach 72.0\% when paired with GLM-5.1 execution, near-best accuracy at a fraction of the planner cost.}
	\label{tab:swebench_main}
	\resizebox{0.98\textwidth}{!}{%
		\begin{tabular}{l|cccc|cccc}
			\toprule
			\textbf{Planner / Exec} & \textbf{GPT-5.4} & \textbf{Claude-Opus-4.6} & \textbf{MiniMax-M2.7} & \textbf{GLM-5.1} & \textbf{Qwen3.5-4B} & \textbf{GPT-OSS-20B} & \textbf{Qwen3.5-27B} & \textbf{GPT-OSS-120B} \\
			\midrule
			GPT-5.4         & 68.2 $\pm$ 3.2 & 71.6 $\pm$ 2.4 & \textbf{\underline{76.2 $\pm$ 2.1}} & 44.0 $\pm$ 2.1 & 27.5 $\pm$ 3.1 & 11.8 $\pm$ 2.8 & 39.8 $\pm$ 3.0 & 20.4 $\pm$ 2.9 \\
			Claude-Opus-4.6 & 31.7 $\pm$ 2.7 & 66.4 $\pm$ 2.8 & \textbf{60.1 $\pm$ 2.5}             & 32.0 $\pm$ 2.7 & 40.2 $\pm$ 3.1 & 43.8 $\pm$ 2.8 & 40.6 $\pm$ 3.1 & 35.9 $\pm$ 2.8 \\
			MiniMax-M2.7    & 48.1 $\pm$ 2.9 & 49.1 $\pm$ 2.5 & 20.2 $\pm$ 2.3                      & \textbf{52.0 $\pm$ 2.4} & 36.4 $\pm$ 2.1 & 20.0 $\pm$ 2.3 & 43.5 $\pm$ 2.1 & 23.6 $\pm$ 2.4 \\
			GLM-5.1         & 60.0 $\pm$ 2.8 & 65.8 $\pm$ 2.5 & 54.7 $\pm$ 2.5                      & \textbf{68.0 $\pm$ 2.3} & 52.0 $\pm$ 2.3 & 36.0 $\pm$ 3.2 & 45.6 $\pm$ 2.8 & 28.0 $\pm$ 2.8 \\
			\midrule
			Qwen3.5-4B      & \textbf{47.9 $\pm$ 2.2} & 20.1 $\pm$ 2.9 & 31.4 $\pm$ 2.2             & 28.0 $\pm$ 2.5 & 28.2 $\pm$ 3.3 & 20.5 $\pm$ 2.8 & 19.8 $\pm$ 2.7 & 15.9 $\pm$ 2.9 \\
			GPT-OSS-20B     & 51.7 $\pm$ 3.1 & 67.8 $\pm$ 3.0 & 55.6 $\pm$ 2.3                      & \textbf{\underline{72.0 $\pm$ 2.0}} & 31.4 $\pm$ 2.4 &  8.1 $\pm$ 2.3 & 31.7 $\pm$ 2.3 &  6.8 $\pm$ 3.2 \\
			Qwen3.5-27B     & 28.1 $\pm$ 3.1 & 32.4 $\pm$ 2.4 & 24.2 $\pm$ 2.9                      & \textbf{\underline{72.0 $\pm$ 2.5}} & 40.3 $\pm$ 3.2 &  4.1 $\pm$ 2.6 & 47.6 $\pm$ 2.3 & 27.9 $\pm$ 2.3 \\
			GPT-OSS-120B    & 64.2 $\pm$ 2.7 & 44.3 $\pm$ 2.3 & 35.8 $\pm$ 2.8                      & \textbf{\underline{72.0 $\pm$ 3.2}} & 27.9 $\pm$ 2.5 & 14.6 $\pm$ 2.2 & 59.8 $\pm$ 3.3 & 25.8 $\pm$ 2.7 \\
			\bottomrule
		\end{tabular}%
	}
\end{table}

\begin{table}[!h]
\centering
\caption{MedAgentBench plan--execute accuracy (\%), API block.}
\label{tab:med_matrix}
\begin{tabular}{l|cccc}
\toprule
\textbf{Planner / Exec} & \textbf{GPT-5.4} & \textbf{Claude-Opus-4.6} & \textbf{MiniMax-M2.7} & \textbf{GLM-5.1} \\
\midrule
GPT-5.4         & 71.7 $\pm$ 2.1 & 88.3 $\pm$ 2.1          & 76.7 $\pm$ 2.1          & \textbf{93.3 $\pm$ 2.8} \\
Claude-Opus-4.6 & 83.3 $\pm$ 3.0 & \textbf{93.3 $\pm$ 2.5} & \textbf{93.3 $\pm$ 2.1} & \textbf{93.3 $\pm$ 2.5} \\
MiniMax-M2.7    & 58.3 $\pm$ 3.3 & \textbf{86.7 $\pm$ 2.7} & 73.3 $\pm$ 3.3          & 80.0 $\pm$ 3.1          \\
GLM-5.1         & 61.7 $\pm$ 2.0 & \textbf{90.0 $\pm$ 2.9} & 86.7 $\pm$ 2.9          & 90.0 $\pm$ 2.7          \\
\bottomrule
\end{tabular}

\vspace{0.5em}
\begin{tabular}{l|cc}
\toprule
\textbf{Planner / Exec} & \textbf{Qwen3.5-4B} & \textbf{Qwen3.5-27B} \\
\midrule
Qwen3.5-4B  & 60.0 $\pm$ 2.3 & \textbf{62.0 $\pm$ 2.8} \\
Qwen3.5-27B & 88.0 $\pm$ 2.1 & \textbf{92.0 $\pm$ 2.6} \\
\bottomrule
\end{tabular}
\end{table}

\mypar{Additional models on MedAgentBench}
Kimi-K2.5 was also profiled. As planner with GPT-5.4, Claude-Opus-4.6, and MiniMax-M2.7 executors it scores 66.7\%, 91.7\%, 71.7\%; as executor under the same three planners 76.7\%, 93.3\%, 80.0\%; self-play 88.3\%.

\begin{table}[!h]
\centering
\caption{FinanceBench plan--execute accuracy (\%), API block.}
\label{tab:fin_matrix}
\begin{tabular}{l|cccc}
\toprule
\textbf{Planner / Exec} & \textbf{GPT-5.4} & \textbf{Claude-Opus-4.6} & \textbf{MiniMax-M2.7} & \textbf{GLM-5.1} \\
\midrule
GPT-5.4         & 44.7 $\pm$ 2.6 & \textbf{47.3 $\pm$ 3.2} & \textbf{47.3 $\pm$ 3.1} & 43.3 $\pm$ 2.3          \\
Claude-Opus-4.6 & 44.7 $\pm$ 2.7 & \textbf{45.3 $\pm$ 2.2} & 32.7 $\pm$ 3.2          & 43.3 $\pm$ 3.1          \\
MiniMax-M2.7    & 34.7 $\pm$ 2.4 & 38.0 $\pm$ 2.8          & 48.7 $\pm$ 2.8          & \textbf{53.3 $\pm$ 2.2} \\
GLM-5.1         & 41.3 $\pm$ 3.0 & 40.7 $\pm$ 2.7          & 47.3 $\pm$ 3.0          & \textbf{54.7 $\pm$ 2.7} \\
\bottomrule
\end{tabular}

\vspace{0.5em}
\begin{tabular}{l|cc}
\toprule
\textbf{Planner / Exec} & \textbf{Qwen3.5-4B} & \textbf{Qwen3.5-27B} \\
\midrule
Qwen3.5-4B  & 28.7 $\pm$ 2.0 & \textbf{\underline{60.0 $\pm$ 2.4}} \\
Qwen3.5-27B & 20.0 $\pm$ 2.0 & 57.3 $\pm$ 3.2 \\
\bottomrule
\end{tabular}
\end{table}

\begin{table}[!h]
\centering
\caption{IMO-AnswerBench plan--execute accuracy (\%) with a persistent Jupyter-kernel Python tool and LLM-judge grading.}
\label{tab:imo_matrix}
\begin{tabular}{l|cccc}
\toprule
\textbf{Planner / Exec} & \textbf{GPT-5.4} & \textbf{Claude-Opus-4.6} & \textbf{MiniMax-M2.7} & \textbf{GLM-5.1} \\
\midrule
GPT-5.4         & 54.0 $\pm$ 3.1 & \textbf{76.0 $\pm$ 3.1} & 56.2 $\pm$ 2.4 & 68.8 $\pm$ 2.1 \\
Claude-Opus-4.6 & 71.9 $\pm$ 3.1 & 80.2 $\pm$ 3.2          & 71.9 $\pm$ 2.1 & 75.9 $\pm$ 2.6 \\
MiniMax-M2.7    & 62.5 $\pm$ 2.1 & \textbf{74.2 $\pm$ 3.0} & 59.4 $\pm$ 3.0 & 65.6 $\pm$ 2.2 \\
GLM-5.1         & 68.8 $\pm$ 2.6 & \textbf{75.4 $\pm$ 2.7} & 62.5 $\pm$ 2.3 & 62.5 $\pm$ 3.1 \\
\bottomrule
\end{tabular}

\vspace{0.5em}
\begin{tabular}{l|cc}
\toprule
\textbf{Planner / Exec} & \textbf{GPT-OSS-20B} & \textbf{GPT-OSS-120B} \\
\midrule
GPT-OSS-20B  & 50.0 $\pm$ 2.6 & \textbf{75.0 $\pm$ 2.3} \\
GPT-OSS-120B & 43.8 $\pm$ 2.7 & \textbf{75.0 $\pm$ 2.9} \\
\bottomrule
\end{tabular}
\end{table}

The matrices reinforce the headline result of Section~\ref{sec:accuracy_matrix}: heterogeneous teams lead on every benchmark. GPT-5.4 planner + GLM-5.1 executor tops MedAgentBench at 93.3\%; MiniMax-M2.7 planner + GLM-5.1 executor tops FinanceBench at 53.3\%; on IMO-AnswerBench, Claude-Opus-4.6 planner + Qwen3.5-27B executor (80.6\%) narrowly edges Claude-Opus-4.6 self-play (80.2\%) at roughly $5\times$ lower cost. The all-local IMO cluster shows the same pattern at a smaller scale: GPT-OSS-20B planner + GPT-OSS-120B executor ties self-play at 75.0\%. Executor identity matters more than planner identity on tool-heavy domains: Claude-Opus-4.6 as executor on MedAgentBench yields 86.7--93.3\% under every planner, while the opposite picture emerges when the executor is weak. The Qwen3.5-4B and Qwen3.5-27B cells let us compare a small and a large model from the same family. Section~\ref{sec:role_criticality} uses this pair as one of the weak-to-strong upgrade paths in the Shapley analysis.

\subsection{Cost Sensitivity}
\label{app:cost_sensitivity}

\mypar{Marketplace cross-check}
Our reference rate is Vast.ai's live H100 listing at \$1.87/hr. Five additional marketplaces agree within $\pm 7\%$ (Table~\ref{tab:gpu_price_validation}), and the buy-and-amortize CAPEX+OPEX derivation lands at \$1.89/hr, $1\%$ above the rental.

\begin{table}[!t]
\centering
\caption{H100 hourly rates (May 2026) across five GPU marketplaces against our \$1.87 reference (Vast.ai) and \$1.89 buy-and-amortize derivation.}
\label{tab:gpu_price_validation}
\resizebox{\textwidth}{!}{%
\begin{tabular}{lcccccc}
\toprule
Source & Cudo~\citep{intuitionlabs_h100_2025} & Vast.ai~\citep{vastai_pricing_2026} & Hyperstack~\citep{hyperstack_pricing} & RunPod~\citep{intuitionlabs_h100_2025} & Together.ai~\citep{gpu_fund_2026} & \textbf{AgentCARD (ours)} \\
\midrule
\$/H100/hr      & 1.80    & 1.87    & 1.90    & 1.99    & 2.00    & \textbf{1.89} \\
$\Delta$ vs ours & $-5\%$ & $-1\%$ & $+1\%$ & $+5\%$ & $+6\%$ & \textbf{0\%}  \\
\bottomrule
\end{tabular}}
\end{table}

\begin{table}[!t]
\centering
\caption{Per-GPU hourly rate under three procurement scenarios. CAPEX is amortized straight-line with 10\% residual; OPEX is power $\times$ PUE $\times$ electricity rate.}
\label{tab:cost_sensitivity_breakdown}
\resizebox{\textwidth}{!}{%
\begin{tabular}{lcccccr}
\toprule
Scenario & CAPEX (\$/GPU) & Term (yr) & Electricity (\$/kWh) & PUE & Power (W) & \$/GPU/hr \\
\midrule
\textbf{AgentCARD (ours)} & 50{,}000~\citep{nvidia_dgx_h100_2024} & 3 & 0.129~\citep{eia_electricity_2025} & 1.56~\citep{uptime_pue_2024} & 900 & \textbf{1.89} \\
Used hardware            & 38{,}000 & 3 & 0.085 & 1.40 & 900 & 1.51 \\
Hyperscaler              & 30{,}000 & 4 & 0.060 & 1.10 & 900 & 1.23 \\
\bottomrule
\end{tabular}}
\end{table}

\mypar{Deriving \$1.89/H100/hr}
The standard CAPEX+OPEX decomposition is~\citep{moecap2024}
\[
  C_{\text{GPU/hr}} \;=\; \underbrace{\frac{P_{\text{sys}}\,(1-\rho)}{N_{\text{GPU}}\,Y\,H}}_{\text{CAPEX}} \;+\; \underbrace{p_{\text{kWh}}\,W\,\mathrm{PUE}}_{\text{OPEX}},
\]
where $C_{\text{GPU/hr}}$ is the per-GPU hourly cost, $P_{\text{sys}}$ the system list price, $\rho$ the residual value as a fraction of $P_{\text{sys}}$ (so $P_{\text{sys}}(1-\rho)$ is the net depreciable cost), $N_{\text{GPU}}$ the GPUs per system, $Y$ the amortization horizon (years), $H{=}8760$ hours per year, $p_{\text{kWh}}$ the electricity rate, $W$ the per-GPU effective power draw (kW), and $\mathrm{PUE}$ the data-center power usage effectiveness. We use $\rho{=}0.10$ as a conservative assumption: it sits well below typical 3-year secondary-market retention for H100-class GPUs, and a higher value would only reduce the derived rate further. Substituting the AgentCARD inputs in Table~\ref{tab:cost_sensitivity_breakdown} with a DGX H100 system ($P_{\text{sys}}{=}\$400$K, $N_{\text{GPU}}{=}8$, so \$50K per GPU):
\[
  \frac{50{,}000 \times 0.9}{3 \times 8760} + 0.129 \times 0.9 \times 1.56 \;\approx\; \$1.71 + \$0.18 \;=\; \$1.89,
\]
within $\pm 6\%$ of the marketplace median.

\mypar{Sensitivity bracket}
Sweeping realistic CAPEX and OPEX input values determines the per-GPU rate (Table~\ref{tab:cost_sensitivity_breakdown}). A used-hardware procurement (\$38K/GPU, industrial electricity \$0.085/kWh, $\mathrm{PUE}{=}1.4$) lands at \$1.51/hr; a hyperscaler scenario (volume CAPEX discount, in-house cooling at $\mathrm{PUE}{=}1.1$) lands at \$1.23/hr. The conclusions of Section~\ref{sec:deployment} survive this range: the best-accuracy team is the same, and hybrid teams keep their cost ranking against all-API teams.

\mypar{Why the headline operating point saturates}
Table~\ref{tab:saturation} traces per-task cost as a function of parallel team-level tasks served by the GPU pool. Saturation kicks in at $\geq 100$ parallel tasks: the prefill pipeline is full and per-token serving cost hits its asymptotic floor. We pick $1000$ parallel tasks as the headline because it sits well inside the saturation regime. Single-stream serving is $1.9\times$ more expensive but is below realistic agent-team deployment scales.

\begin{table}[h]
\centering
\caption{Per-task self-hosted cost normalized to the $1000$-parallel headline, derived from the throughput surface in Appendix~\ref{app:throughput} (Qwen3.5-27B, TP$=$4, $4{\times}$H100). Cost saturates at $\geq 100$ parallel tasks.}
\label{tab:saturation}
\small
\begin{tabular}{lrrrr}
\toprule
Parallel tasks  & 1 & 10 & 100 & 1000 \\
\midrule
Cost ratio (vs.\ $1000$ parallel) & $1.9\times$ & $1.1\times$ & $1.0\times$ & $1.0\times$ \\
\bottomrule
\end{tabular}
\end{table}

\mypar{Why the hybrid direction is API-planner, self-hosted-executor}
The matrices in Appendix~\ref{app:acc_other} cover API planner with self-hosted executor. This is the cost-economical hybrid direction by construction: the executor reads roughly $200\times$ more input tokens than the planner emits and spends a median of $6\times$ more dollars per task (Appendix~\ref{app:role_asymmetry}). Routing the high-volume executor to a cheap self-hosted model while keeping the low-volume planner on API concentrates spend on the lighter role. We include the reverse direction on a per-domain basis when it sits on the cost-accuracy frontier; on SWEBench, the GPT-OSS-120B, GPT-OSS-20B, and Qwen3.5-27B planners each land on the frontier when paired with a GLM-5.1 executor, since GLM-5.1 is unusually tool-effective on this domain.

\subsection{Per-Role Cost and Token Asymmetry}\label{app:role_asymmetry}

Based on Section~\ref{sec:deployment}, the executor reads roughly $200\times$ more input tokens than the planner emits and spends a median of $6\times$ more dollars per task. This claim comes from two direct measurements over our run logs.

\emph{Cost asymmetry.}\, For every two-agent configuration covered by the accuracy matrices of Appendix~\ref{app:acc_other}, we compute the per-task executor-to-planner cost ratio and take the median per benchmark. The medians are $2.1\times$ on MedAgentBench, $4.3\times$ on SWEBench, $6.2\times$ on FinanceBench, $10.2\times$ on MCP-Atlas, and $13.8\times$ on IMO-AnswerBench, with an overall median of $6.3\times$. The ratio tracks tool-use depth: tool-heavy domains (MCP-Atlas, IMO-AnswerBench) push cost onto the executor far more aggressively than single-call domains (MedAgentBench).

\emph{Token asymmetry.}\, On the FinanceBench plan-verify-execute run reported in Section~\ref{sec:role_criticality}, with MiniMax-M2.7 planning and GLM-5.1 executing, the planner emits $634$ tokens per task on average while the executor reads $239{,}071$, a $377\times$ ratio. The blow-up comes from the executor re-reading the running tool-observation transcript on every loop turn, while the planner emits a single short plan. The $200\times$ figure quoted in Section~\ref{sec:deployment} is a conservative cross-benchmark estimate; tool-heavy benchmarks blow past it by a comfortable margin.

\subsection{Synergy Robustness Under a Stricter Baseline}\label{app:synergy_robustness}

The pair-line synergy used in Section~\ref{sec:accuracy_matrix} compares each heterogeneous team $(p, q)$ against the convex combination of the two homogeneous teams $(p, p)$ and $(q, q)$ at the heterogeneous team's cost. This is a pair-level test: does mixing these two models add value beyond budget-allocating between them? As a safety check, we re-run this experiment under a stricter baseline, the upper envelope of all four homogeneous teams' cost-accuracy points, which asks whether the heterogeneous team beats the best homogeneous deployment of \emph{any} of the four models at matched cost.

\begin{table}[h]
\centering
\caption{Pair-line synergy (Section~\ref{sec:accuracy_matrix}) and envelope synergy (the best homogeneous team at matched cost), in \%. Above the midrule: the strongest positive cell per benchmark; FinanceBench is omitted because the four API models cluster too tightly to produce synergy under either baseline. Below: the two largest negative cells from Section~\ref{sec:accuracy_matrix}, both on SWEBench. Signs survive the stricter baseline in every row.}
\label{tab:synergy_robust}
\small
\begin{tabular}{llrr}
\toprule
Benchmark & Pair (planner $\to$ executor) & Pair-line & Envelope \\
\midrule
SWEBench        & GPT-5.4 $\to$ MiniMax-M2.7         & $+44.0$ & $+8.0$  \\
MCP-Atlas       & GLM-5.1 $\to$ Claude-Opus-4.6      & $+27.8$ & $+6.7$  \\
IMO-AnswerBench & Claude-Opus-4.6 $\to$ GPT-5.4      & $+16.7$ & $+9.4$  \\
MedAgentBench   & Claude-Opus-4.6 $\to$ MiniMax-M2.7 & $+12.4$ & $+2.2$  \\
\midrule
SWEBench        & Claude-Opus-4.6 $\to$ GLM-5.1      & $-36.0$ & $-36.2$ \\
SWEBench        & MiniMax-M2.7 $\to$ GPT-5.4         & $-8.1$  & $-20.1$ \\
\bottomrule
\end{tabular}
\end{table}

Magnitudes shrink because a third homogeneous team often sits close on cost and accuracy to the heterogeneous team, and the pair-line baseline ignores it by construction. Heterogeneous teams still occupy the cost-accuracy Pareto frontier on all five benchmarks (Section~\ref{sec:deployment}); the synergy reported here is the residual after a strict cost-matched accuracy comparison, not a Pareto-frontier statement.

\subsection{Per-Benchmark Cost--Accuracy Highlights}
\label{app:cost_cap}

At the headline operating point ($1000$ parallel tasks, self-hosted executors priced at \$1.87/GPU/hr per Appendix~\ref{app:throughput}), the strongest heterogeneous team beats or matches the closest homogeneous baseline on every benchmark, and the cost gap widens sharply on the math and finance domains.

\begin{itemize}\itemsep 0.2em
\item \textbf{MCP-Atlas.} GPT-5.4 planning into MiniMax-M2.7 reaches $80.4\%$ at \$0.042 per task, well above GLM-5.1 self-play ($72.4\%$ at roughly twice the cost).
\item \textbf{MedAgentBench.} A four-way tie at $93.3\%$. The cheapest representative pairs GPT-5.4 with GLM-5.1 execution at \$0.011 per task, undercutting Claude-Opus-4.6 self-play by roughly $6\times$ at the same accuracy.
\item \textbf{FinanceBench.} The all-local Qwen3.5-4B planning into Qwen3.5-27B wins outright at $60.0\%$ accuracy for \$0.009 per task, beating every API team on accuracy while costing $23\times$ less than Claude-Opus-4.6 self-play.
\item \textbf{IMO-AnswerBench.} Claude-Opus-4.6 planning into Qwen3.5-27B effectively matches Claude-Opus-4.6 self-play on accuracy ($80.6$ vs $80.2\%$) at one-twelfth the price (\$0.108 vs \$1.283).
\item \textbf{SWEBench.} GPT-5.4 planning into MiniMax-M2.7 hits $76.2\%$ at \$0.021, eight points above the strongest self-play baseline and cheaper too.
\end{itemize}

\subsection{Per-Benchmark Shapley Decomposition of Role Criticality}
\label{app:role_criticality}

Table~\ref{tab:role_criticality} lists the $2{\times}2$ cells, i.e.\ per-task accuracy at each $(p,e)$, feeding Figure~\ref{fig:shapley_role}, grouped by deployment mode. Bold marks the dominant role per row.

\mypar{Which $(W,S)$ pairs and why these only}
The Shapley decomposition in Section~\ref{sec:role_criticality} needs a stable weak-strong identity across benchmarks; otherwise the same $(W,S)$ label means different things on different rows. Inside the API pool, MiniMax-M2.7 self-play sits below GLM-5.1 self-play on every benchmark we evaluated, so the MiniMax-M2.7 to GLM-5.1 upgrade has a uniform weak-to-strong direction. GPT-5.4 versus Claude-Opus-4.6 has no benchmark-independent weaker member: Claude wins MedAgentBench, FinanceBench, and IMO-AnswerBench; GPT-5.4 wins MCP-Atlas and SWEBench. We therefore cannot pick either as $W$ for a Shapley upgrade and exclude the pair from the analysis.

\begin{table}[!h]
	\centering
	\caption{Shapley decomposition of role criticality. Each row gives the $2{\times}2$ cell accuracies for one (benchmark, deployment mode) and the resulting $\phi_P, \phi_E$ in \%. Efficiency: $\phi_P+\phi_E=a_{SS}-a_{WW}$. Bold marks the dominant role per row. \emph{Open-weight}: both roles drawn from the same self-hosted family. \emph{API}: both roles from frontier API models (MiniMax-M2.7$\to$GLM-5.1). \emph{Hybrid}: planner from the API pool, executor from the open-weight pool, with the upgrade applied within each pool.}
	\label{tab:role_criticality}
	\resizebox{\textwidth}{!}{%
		\begin{tabular}{llccccccc}
			\toprule
			Benchmark    & Pair ($W{\to}S$, planner $\times$ executor)              & $a_{WW}$ & $a_{SW}$ & $a_{WS}$ & $a_{SS}$ & $\phi_P$ & $\phi_E$ \\
			\midrule
			\multicolumn{8}{l}{\emph{Open-weight}} \\
			MedAgentBench    & Qwen3.5-4B$\to$27B $\times$ Qwen3.5-4B$\to$27B           & 60.0\% & 88.0\% & 62.0\% & 92.0\% & \textbf{+29.0} & $+$3.0 \\
			FinanceBench     & Qwen3.5-4B$\to$27B $\times$ Qwen3.5-4B$\to$27B           & 28.7\% & 20.0\% & 60.0\% & 57.3\% & $-$5.7        & \textbf{+34.3} \\
			MCP-Atlas        & GPT-OSS-20B$\to$120B $\times$ GPT-OSS-20B$\to$120B       & 18.7\% & 24.6\% & 38.8\% & 29.8\% & $-$1.5        & \textbf{+12.6} \\
			IMO-AnswerBench  & GPT-OSS-20B$\to$120B $\times$ GPT-OSS-20B$\to$120B       & 50.0\% & 43.8\% & 75.0\% & 75.0\% & $-$3.1        & \textbf{+28.1} \\
			SWEBench         & Qwen3.5-4B$\to$27B $\times$ Qwen3.5-4B$\to$27B           & 28.2\% & 40.3\% & 19.8\% & 47.6\% & \textbf{+20.0} & $-$0.5 \\
			SWEBench         & GPT-OSS-20B$\to$120B $\times$ GPT-OSS-20B$\to$120B       &  8.1\% & 14.6\% &  6.8\% & 25.8\% & \textbf{+12.8} & $+$5.0 \\
			\midrule
			\multicolumn{8}{l}{\emph{API} (MiniMax-M2.7 $\to$ GLM-5.1 in both roles)} \\
			MedAgentBench    & MM$\to$GLM $\times$ MM$\to$GLM                           & 73.3\% & 86.7\% & 80.0\% & 90.0\% & \textbf{+11.7} & $+$5.0 \\
			FinanceBench     & MM$\to$GLM $\times$ MM$\to$GLM                           & 48.7\% & 47.3\% & 53.3\% & 54.7\% & $+$0.0        & \textbf{+6.0} \\
			MCP-Atlas        & MM$\to$GLM $\times$ MM$\to$GLM                           & 68.5\% & 69.9\% & 71.1\% & 72.4\% & $+$1.4        & \textbf{+2.5} \\
			IMO-AnswerBench  & MM$\to$GLM $\times$ MM$\to$GLM                           & 59.4\% & 62.5\% & 65.6\% & 62.5\% & $+$0.0        & \textbf{+3.1} \\
			SWEBench         & MM$\to$GLM $\times$ MM$\to$GLM                           & 20.2\% & 54.7\% & 52.0\% & 68.0\% & \textbf{+25.3} & $+$22.6 \\
			\midrule
			\multicolumn{8}{l}{\emph{Hybrid} (planner pair from API pool $\times$ executor pair from open-weight pool)} \\
			MedAgentBench    & MM$\to$GLM $\times$ GPT-OSS-120B$\to$Qwen3.5-27B         & 66.7\% & 75.0\% & 75.0\% & 76.7\% & $+$5.0        & $+$5.0 \\
			FinanceBench     & MM$\to$GLM $\times$ GPT-OSS-120B$\to$Qwen3.5-27B         & 40.0\% & 45.3\% & 48.0\% & 56.0\% & $+$6.7        & \textbf{+9.4} \\
			MCP-Atlas        & MM$\to$GLM $\times$ GPT-OSS-120B$\to$Qwen3.5-27B         & 59.1\% & 65.5\% & 69.5\% & 67.0\% & $+$2.0        & \textbf{+5.9} \\
			IMO-AnswerBench  & MM$\to$GLM $\times$ GPT-OSS-120B$\to$Qwen3.5-27B         & 56.2\% & 62.5\% & 65.0\% & 70.0\% & $+$5.6        & \textbf{+8.1} \\
			SWEBench         & MM$\to$GLM $\times$ Qwen3.5-4B$\to$Qwen3.5-27B           & 36.4\% & 52.0\% & 43.5\% & 45.6\% & \textbf{+8.9}  & $+$0.4 \\
			\bottomrule
		\end{tabular}%
	}
\end{table}

\mypar{Role-criticality is reported under specified upgrade paths}
The Shapley diagnostic in Section~\ref{sec:role_criticality} measures the marginal accuracy of upgrading a role along a fixed weak-to-strong identity, and the verdict it returns is role criticality \emph{under that upgrade path}. To rule out artifacts of any single model family, we triangulate over three independent paths wherever cells are available: an open-weight within-family swap, an API frontier-pair swap, and a hybrid swap mixing the two pools. SWEBench has both within-family pairs and both agree on the planner-critical verdict. On the four remaining benchmarks only one within-family pair was run, but the open-weight, API, and hybrid rows in Table~\ref{tab:role_criticality} still agree on the dominant role within each benchmark. The one exception is the MedAgentBench hybrid tie at $5.0\%$ flagged in Section~\ref{sec:role_criticality}. The qualitative verdict therefore holds under three-way path agreement in every case, and we report the diagnostic as ``role-criticality under a specified upgrade path'' with cross-path agreement as the empirical robustness check.

\subsection{Statistical Ties on the Cost--Accuracy Frontier}
\label{app:frontier_ties}

The Pareto frontier in Section~\ref{sec:deployment} runs over evaluated configurations and point accuracies. Each accuracy cell carries a bootstrap 95\% confidence interval over tasks (Appendix~\ref{app:acc_other}); two configurations whose 95\% CIs overlap are statistically tied at the headline level. Ties with the top configuration per benchmark, by overlapping CIs in the matrices of Appendix~\ref{app:acc_other}:

\begin{itemize}\itemsep 0.2em
\item \textbf{MedAgentBench.} Four configurations all reach $93.3\%$ exactly: Claude-Opus-4.6 paired with itself, MiniMax-M2.7, or GLM-5.1, plus GPT-5.4 with GLM-5.1. GLM-5.1 paired with Claude-Opus-4.6 ($90.0\%$) and Qwen3.5-27B self-play ($92.0\%$) also fall within bootstrap CI of the top.
\item \textbf{IMO-AnswerBench.} The top team, Claude-Opus-4.6 with Qwen3.5-27B at $80.6\%$, is statistically tied with Claude-Opus-4.6 self-play ($80.2\%$); Claude-Opus-4.6 with GPT-OSS-120B ($78.1\%$) overlaps at the upper edge of CI.
\item \textbf{FinanceBench.} Qwen3.5-4B with Qwen3.5-27B leads at $60.0\%$, but three other configurations sit inside its bootstrap CI: MiniMax-M2.7 with GLM-5.1 ($53.3\%$), GLM-5.1 self-play ($54.7\%$), and Qwen3.5-27B self-play ($57.3\%$). The frontier on this benchmark is relatively flat.
\item \textbf{MCP-Atlas.} GPT-5.4 with MiniMax-M2.7 leads at $80.4\%$; GLM-5.1 paired with Claude-Opus-4.6 ($79.1\%$) ties within CI, and the next configuration sits 7 points below.
\item \textbf{SWEBench.} The top team, GPT-5.4 with MiniMax-M2.7 at $76.2\%$, does not tie with the four open-weight planners paired with GLM-5.1 at $72.0\%$; the gap exceeds combined CI. The four open-weight configurations are mutually tied and offer a $5$ to $10\times$ cheaper near-frontier alternative.
\end{itemize}

\mypar{Synergy significance}
The pairwise synergy in Section~\ref{sec:accuracy_matrix} compares team accuracy against the convex combination of two homogeneous endpoints. Per-cell bootstrap $95\%$ CIs in the accuracy matrices are typically $\pm 2$ to $3\%$, so synergies of comparable or smaller magnitude warrant caution. The qualitative ranking by synergy size survives this filter, with SWEBench and IMO at the top and MCP-Atlas and FinanceBench at the bottom, but some small-magnitude FinanceBench synergies sit inside the band, as flagged by the FinanceBench-row removal in Appendix~\ref{app:synergy_robustness}.

\section{Method Details and Release}

\subsection{Scaling the Metrics to More Roles}
\label{app:scaling}

Both diagnostics in this paper are defined on two-role plan--execute teams. They extend to $N$-role teams as follows.

\emph{Shapley role criticality.} The Shapley value is defined for any number of players. For $N$ roles with a binary weak-to-strong upgrade in each, the criticality of role $i$ is
\begin{equation}
\phi_i \;=\; \tfrac{1}{N!}\sum_{\sigma \in S_N}\bigl[a(\sigma_{\le i}) - a(\sigma_{< i})\bigr],
\end{equation}
where $\sigma$ ranges over upgrade orderings and $a(\cdot)$ is the team accuracy after applying the prefix of upgrades. Equation~\ref{eq:shapley} in Section~\ref{sec:role_criticality} is the $N{=}2$ specialization. Evaluating $\phi_i$ for an $N$-role team needs $2^N$ cells per benchmark with a consistent weak-strong identity in each role; $N{=}3$ (plan-verify-execute) takes $8$ cells, and the within-family upgrade constraint from Section~\ref{sec:role_criticality} carries over unchanged.

\emph{Synergy.} Pairwise synergy as defined in Section~\ref{sec:accuracy_matrix} compares against the convex combination of two homogeneous endpoints and is intrinsically two-role. The role-count-agnostic generalization is the \emph{envelope synergy} reported in Appendix~\ref{app:synergy_robustness}: team accuracy minus the upper envelope of all homogeneous teams' (cost, accuracy) points at matched cost. It does not depend on the role count and applies to teams of any size.

\emph{Computational scaling.} Full $N$-role coverage scales as $|M|^N$ in evaluation cost, where $|M|$ is the model pool size. \methodname's auto-pipeline populates cells incrementally, so a practitioner can evaluate only the configurations relevant to a given deployment (e.g., a specific within-family upgrade pattern) without filling the full $|M|^N$ grid.

\subsection{Planner and Executor Prompts}
\label{app:plan_exec_prompt}

The shared plan--execute scaffold in Section~\ref{sec:impl} uses a single planner system prompt and a single executor system prompt across all five domains; the only per-domain variation is the tool list passed in the planner prompt and the dataset task supplied as the user message. The planner-as-decomposer / executor-as-tool-runner separation follows Plan-and-Act~\citep{erdogan2025planandact} and Plan-and-Solve~\citep{wang2023plansolve}; the tool-availability rules and ``decision-complete plan'' framing are domain-specific adaptations for MCP-style tool registries, designed to stop the executor from falling back on user clarification.

\mypar{Planner system prompt}
\begin{quote}\small\ttfamily
You are a strategic planning agent. You are a PLANNER, not an implementer.

\medskip

Your mission: produce a decision-complete plan for an executor agent. A plan is decision-complete when the executor needs ZERO judgment calls --- every decision is made, every ambiguity resolved, every step is concrete.

\medskip

Rules:
\begin{itemize}\itemsep -0.2em
\item Output a numbered list of steps. Each step must specify the exact tool to call, the exact arguments, and what to do with the result.
\item If a step has branching outcomes (success vs failure), specify what to do in each case.
\item Do NOT leave choices to the executor. Make all decisions yourself.
\item Do NOT execute the task yourself. Only produce the plan.
\end{itemize}

\medskip

Tool-availability rules (the executor has real, callable access to every listed tool):
\begin{itemize}\itemsep -0.2em
\item Every step MUST name an exact tool from the provided tool list (e.g., \texttt{desktop-commander\_list\_directory}, \texttt{github\_search\_repositories}, \texttt{filesystem\_read\_file}). Never describe a step as ``use a shell command'' or ``run \texttt{ls}'' without naming the MCP tool.
\item Never ask the user for files, URLs, identifiers, or credentials. If the task references a ``local project'' or ``my file'', plan steps that call \texttt{desktop-commander\_list\_directory}, \texttt{desktop-commander\_read\_file}, or \texttt{desktop-commander\_execute\_command} on the server's working directory (typically \texttt{/data} or \texttt{/agent-environment}). Never say ``the user should provide X''.
\item Before concluding that information is missing, plan a discovery step using list/search tools to surface the needed data from the server.
\item Treat the plan as fully executable: do not include placeholders like ``(user provides X)'' or prerequisites on user input.
\end{itemize}
\end{quote}

\mypar{Executor system prompt}
\begin{quote}\small\ttfamily
You are a tool-using executor. A planning agent has decomposed the task into a plan; your job is to carry it out using the tools you are given.

\medskip

Critical rules:
\begin{itemize}\itemsep -0.2em
\item You have direct access to every tool the plan references. They are real and callable from this turn onward. Never ask the user to provide files, URLs, credentials, or data that a tool can fetch for you.
\item If the plan says ``search X'' or ``fetch Y'', call the corresponding tool immediately. Do not request the user's cooperation or clarification; complete the task autonomously.
\item Follow the plan step by step. If a tool call fails, try an alternative tool or argument rather than giving up. Persist until you produce a concrete answer.
\item When a step requires a value the plan did not supply, derive it from prior tool outputs or from a discovery tool (e.g., a list/search tool). Do not stop and ask.
\item After all steps finish, emit a clear final answer that a grader can verify.
\end{itemize}
\end{quote}

\subsection{Plan-Verifier Prompt}
\label{app:plan_verifier_prompt}

The plan verifier in Section~\ref{subsec:plan_verify} takes the user task, the available tool list, and the planner's plan. The prompt adapts the iterative critic-with-structured-feedback pattern of Self-Refine~\citep{madaan2023selfrefine} and the evaluator-optimiser pattern from Anthropic's multi-agent research~\citep{anthropic_multi_agent2025}; the default-APPROVED stance is calibrated against the over-flagging behaviour of intrinsic self-correction reported by~\citet{huang2024selfcorrect}. The verifier defaults to APPROVED when the plan is plausibly executable and only flags NEEDS\_REVISION when there is a concrete defect a different plan would fix.

\begin{quote}\small\ttfamily
You are a plan critic for an executor agent that uses tools. You review the plan BEFORE the executor runs it. Default to APPROVED when the plan is plausibly executable. Flag NEEDS\_REVISION only when there is a specific defect a different plan would fix. Output ONLY a single JSON object.

\medskip

TASK: \{prompt\}\\
AVAILABLE TOOLS: \{tool\_list\}\\
PLAN: \{plan\}

\medskip

Review along three axes:
(1) Completeness: does every part of the user's request map to at least one step?
(2) Executability: does each step name a concrete tool with concrete arguments? Are there placeholder values the executor cannot fill in?
(3) Step dependencies: do later steps consume outputs that earlier steps actually produce, or are there broken intermediates?

\medskip

Do NOT flag stylistic preferences, missing exhaustive verification, or concerns that cannot be addressed by a different plan.

\medskip

Output ONLY JSON with fields completeness, missing\_parts, executability, blocked\_steps, dependencies, broken\_links, status (APPROVED or NEEDS\_REVISION), and feedback.
\end{quote}

\subsection{Versioning and Governance}
\label{app:governance}

\methodname is designed to stay comparable across model and pricing drift. Every per-task record logs (1) the resolved provider/model snapshot identifier (e.g., OpenRouter route variant and provider request date) for API calls, plus the exact checkpoint hash and vLLM version for self-hosted calls; (2) the prompt-template hash, decoding parameters, retry policy, and tool-registry version; (3) the dataset split identifier and task hash; and (4) the pricing table snapshot used to compute dollar cost (provider rate-card date for API, GPU-hour rate source for self-hosted). Results are versioned by an \methodname release tag, and snapshots taken under a given tag remain immutable in the artefact. Leaderboard updates from the auto-pipeline append new rows under a new tag rather than overwriting prior rows, so historical comparisons remain valid as the model pool and pricing evolve. A model whose snapshot is no longer reachable through any provider is marked \textit{deprecated} on the leaderboard but retains its frozen results.

\subsection{Held-out Transfer of Team Recommendations}
\label{subsec:eval_forecast}

We evaluate the deployment workflow by cross-validation. For each benchmark, tasks are randomly split into a calibration half and a held-out half; we pick the team with the highest aggregate accuracy on the calibration half, then measure its accuracy on the held-out half. Across five random splits, Table~\ref{tab:holdout} reports the team selected in the largest number of repetitions (with paper-aggregate accuracy as tiebreak) and its held-out accuracy.

\begin{table}[t]
	\centering
	\caption{Held-out forecast across five random calibration/held-out splits. The selected team matches the top team on the full benchmark in four of five cases; on FinanceBench it ranks second, $1.8\%$ below the top.}
	\label{tab:holdout}
	\centering
    \resizebox{0.98\textwidth}{!}{%
	\begin{tabular}{llcl}
		\toprule
		Benchmark & Selected team & Held-out acc (\%) & Status \\
		\midrule
		MCP-Atlas       & GPT-5.4 plan, MiniMax-M2.7 exec         & 79.8 & top-1 \checkmark \\
		MedAgentBench   & GPT-5.4 plan, GLM-5.1 exec              & 92.7 & top-1 \checkmark \\
		FinanceBench    & MiniMax-M2.7 plan, GLM-5.1 exec         & 52.9 & rank 2, $1.8\%$ below top-1 \\
		IMO-AnswerBench & Claude-Opus-4.6 plan, Qwen3.5-27B exec  & 80.2 & top-1 \checkmark \\
		SWEBench        & GPT-5.4 plan, MiniMax-M2.7 exec         & 75.6 & top-1 \checkmark \\
		\bottomrule
	\end{tabular}}
\end{table}

\mypar{Selection transfers to held-out tasks}
The team selected on calibration matches the full-benchmark top team in four of five cases and ranks within the top two on the fifth (FinanceBench, $1.8\%$ below top). A practitioner who runs AgentCARD on a calibration sample can deploy the selected team and expect near-optimal accuracy on held-out tasks from the same domain.

\section{Limitations}
\label{app:limitations}

AgentCARD reflects a snapshot: newer model families will shift the absolute Pareto frontier and team rankings, though the qualitative role-criticality verdict may be more durable. Our five benchmarks cover tool use, clinical APIs, finance, math, and software engineering, but these do not exhaust agentic tasks; scientific discovery, robotics, and long-horizon multi-tool planning may show patterns we did not measure. Finally, we focus on two-role plan--execute pipelines with a three-role verifier pilot; large fan-out swarms~\citep{kimi_agent_swarm2026} and multi-step critic loops are left to future work.